\documentstyle[11pt,newpasp,twoside,epsf]{article}
\def\edcomment#1{\iffalse\marginpar{\raggedright\sl#1\/}\else\relax\fi}
\reversemarginpar

\begin{document}
\title{NLTE Model Atmospheres \\ for Central Stars of Planetary Nebulae}
\author{Thomas Rauch, Jochen L.\,Deetjen, Stefan Dreizler, Klaus Werner}
\affil{Institut f\"ur Astronomie und Astrophysik, T\"ubingen, Germany}

\begin{abstract}
  Present observational techniques provide stellar spectra with high resolution
at a high signal-to-noise ratio over the complete wavelength range -- from the
far infrared to the X-ray.

  NLTE effects are particularly important for hot stars, hence the use of reliable NLTE
stellar model atmosphere fluxes is required for an adequate spectral analysis.

  State-of-the-art\, NLTE\, model atmospheres include the metal-line blanketing of
millions of lines of all elements from hydrogen up to the iron-group elements 
and thus permit precise analyses of extremely hot compact stars, e.g. central 
stars of planetary nebulae, PG\,1159 stars, white dwarfs, and neutron stars.  
Their careful spectroscopic study is of great interest in several branches of
modern astrophysics, e.g.\,stellar and galactic evolution, and interstellar matter.
\end{abstract}

\section{Introduction}
During their evolution, the more massive post-AGB stars can reach extremely high
effective temperatures: Up to about 700\,kK are predicted by Paczynski (1970)
for a star with a remnant mass of 1.2\,M$_\odot$.
Realistic modeling of the emergent fluxes of these stars requires the consideration
of all elements from hydrogen up to the iron group under NLTE conditions.

\section{NLTE Model Atmospheres}
The model atmospheres (plane-parallel, hydrostatic and radiative equilibrium)
are calculated using the code {\sc PRO2} (Werner 1986). All elements from hydrogen
to the iron group are considered (Rauch 1997, Dreizler \& Werner 1993, Deetjen et al.\,1999).
A grid of model atmosphere fluxes ($T_{\rm eff}$ = 50 -- 1\,000\,kK, $\log g$ = 5 -- 9 (cgs), H -- Ca, solar and halo abundances)
is available at the WWW (http://astro.uni-tuebingen.de/$^\sim$rauch/flux.html).

\section{Impact of light metals (F -- Ca)}
The high-energy model atmosphere fluxes strongly depend on the metal-line blanketing. The
impact of H -- Ca is shown in Fig.\,1 (cf.\,Rauch 1997).
\vspace{2mm}

\begin{minipage}{8cm}
\epsfxsize=8cm
\epsffile{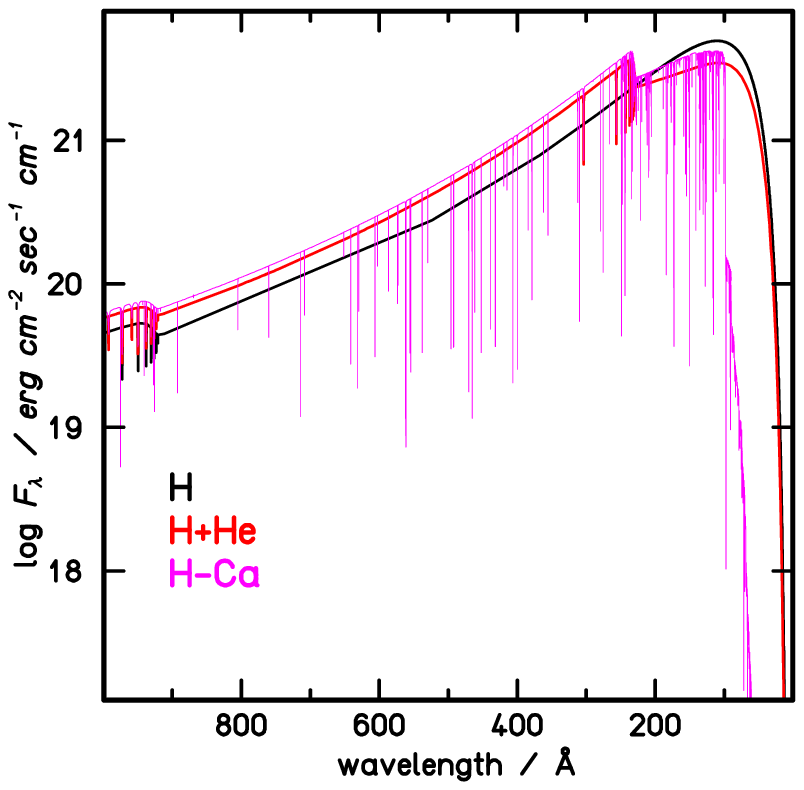}
\end{minipage}\hspace{5mm}
\begin{minipage}{4cm}
Figure 1. Comparison of NLTE model atmosphere fluxes with different elemental
composition at solar abundances ($T_{\rm eff} = 155$\,kK, $\log g = 6.5$). Note
the drastic decrease of the flux level at wavelengths shorter than 100\,\AA\
if the metal-line blanketing of the light metals H-Ca is considered. 
\end{minipage}

\section{Impact of iron-group elements (Sc -- Ni)}
A detailed consideration of all line transitions of the iron-group elements,
like tabulated in Kurucz (1996), is impossible. Thus, we employed an opacity
sampling method in order to account for their absorption cross-sections.
Their cross-sections are calculated with the newly developed Cross-Section Creation Package {\sc CSC}
(http://astro.uni-tuebingen.de/$^\sim$deetjen/csc.html). Radiative und collisional bound-bound 
line cross-sections are calculated from Kurucz's line lists (1996; Fig.\,2).
Radiative und collisional bound-free photoionization cross-sections of Fe are calculated from Opacity Project data 
(Seaton et al.\, 1994).
\vspace{2mm}

\epsfxsize=\textwidth
\epsffile{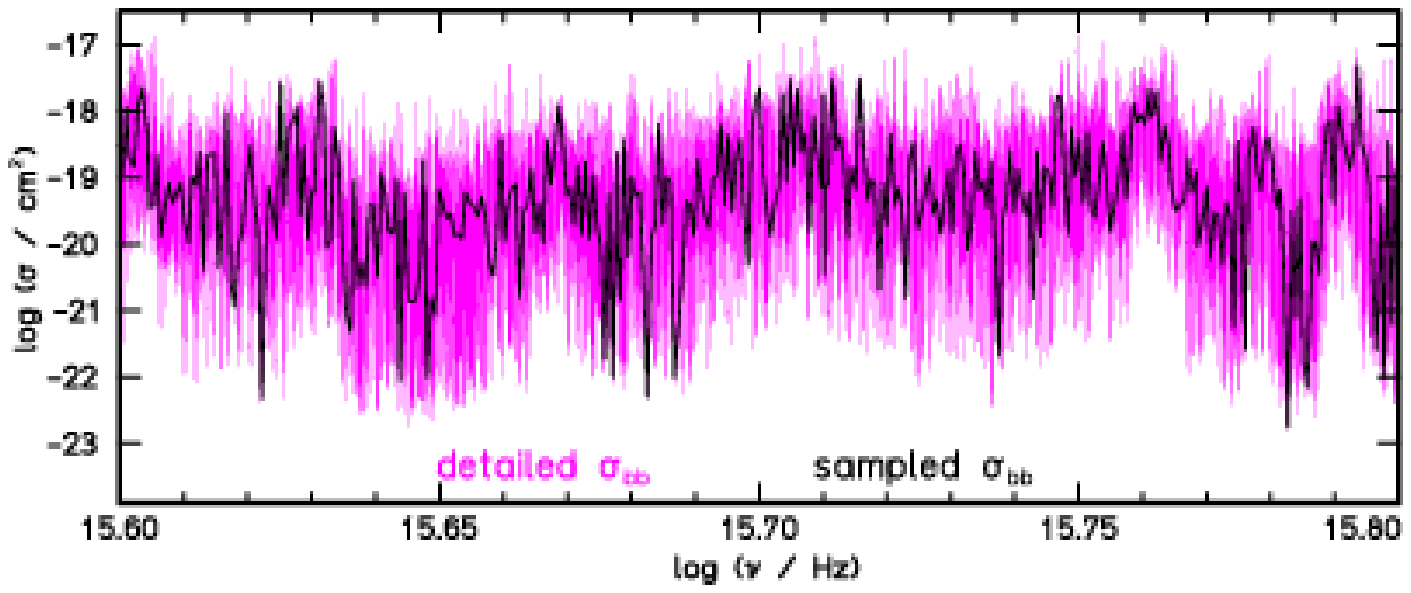}
Figure 2.  Part of the radiative bound-bound cross-section $\sigma_{\rm 4,6}$ (band 6 to band 4) of Fe\,{\sc v} 
at $n_{\rm e} = 10^{16}$\,cm$^{-3}$, considered in detail (520\,360 frequency points within 
$2.0\cdot 10^{15}$\,Hz $\leq \nu \leq 6.8\cdot 10^{15}$\,Hz)
and with our opacity sampling grid (973 frequency points)
\vspace{5mm}

All iron-group elements can be combined in one generic model atom.
Its term scheme is typically divided into seven energy bands (Fig.\,3, Dreizler \& Werner 1993). 
The statistics of a typical generic model atom are summarized in Tab.\,1.
\vspace{2mm}

\begin{minipage}{8cm}
\epsfxsize=8cm
\epsffile{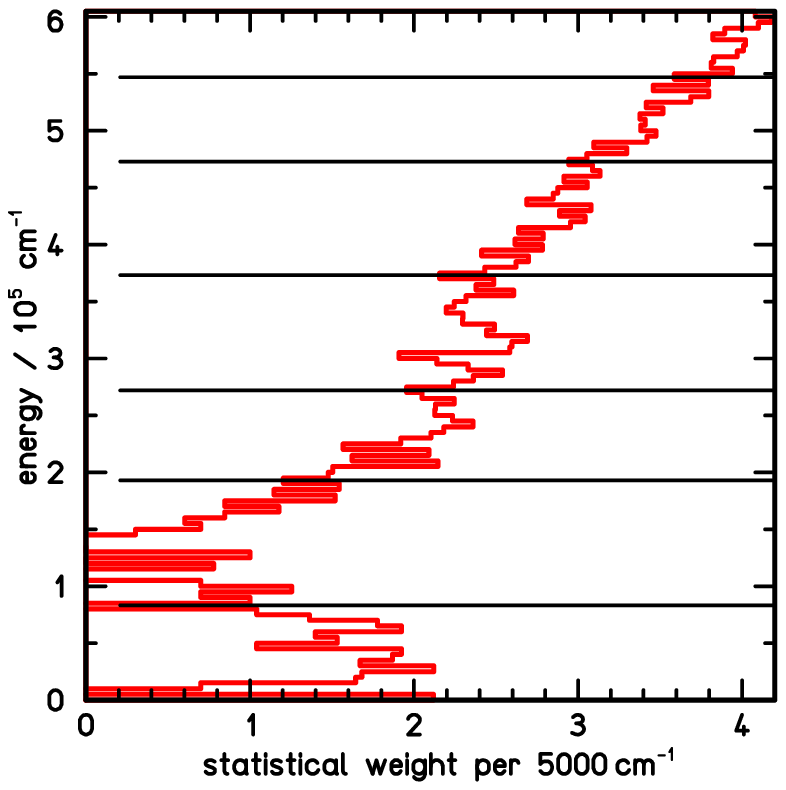}
\end{minipage}\hspace{5mm}
\begin{minipage}{4cm}
Figure 3. Energy band structure of an iron-group model atom
\end{minipage}

\begin{table}\centering
  \caption{Summary of a generic Sc-Ni model atom used in our model atmosphere
    calculations. Numbers in brackets denote individual levels and lines
    used in the statistical NLTE line-blanketing approach}
  \vspace{2mm}
  \begin{tabular}{l r r r r}
    \hline  
    \noalign{\smallskip}
    ion & \multicolumn{2}{c}{NLTE levels} & \multicolumn{2}{c}{line transitions} \\
    \noalign{\smallskip}
    \hline
    \noalign{\smallskip}
    {\sc v   } &  \hbox{}\hspace{5mm}7 & (20\,437)\hspace{5mm}\hbox{} &  \hbox{}\hspace{5mm}26 & (6\,042\,725)  \\
    {\sc vi  } &  \hbox{}\hspace{5mm}7 & (16\,062)\hspace{5mm}\hbox{} &  \hbox{}\hspace{5mm}26 & (4\,784\,314)  \\
    {\sc vii } &  \hbox{}\hspace{5mm}7 & (12\,870)\hspace{5mm}\hbox{} &  \hbox{}\hspace{5mm}26 & (2\,573\,617)  \\
    {\sc viii} &  \hbox{}\hspace{5mm}7 &  (9\,144)\hspace{5mm}\hbox{} &  \hbox{}\hspace{5mm}28 & (3\,229\,141)  \\
    \noalign{\smallskip}
    \hline  
    \noalign{\smallskip}
    total      & \hbox{}\hspace{5mm}28 & (58\,513)\hspace{5mm}\hbox{} & 106 & (16\,656\,797) \\
    \noalign{\smallskip}
    \hline
  \end{tabular}
\end{table}

In order to investigate the impact of the iron-group elements on the emergent flux,
we used a H-Ca trunk model atom (Rauch 1997) and added Sc -- Ni in form of a generic model atom (Tab.\,1), 
including all available (experimental + theoretical) levels and lines from Kurucz's list (1996).
In Fig.\,4 we show the additional impact of the iron group elements.
\vspace{2mm}

\begin{minipage}{8cm}
\epsfxsize=8cm
\epsffile{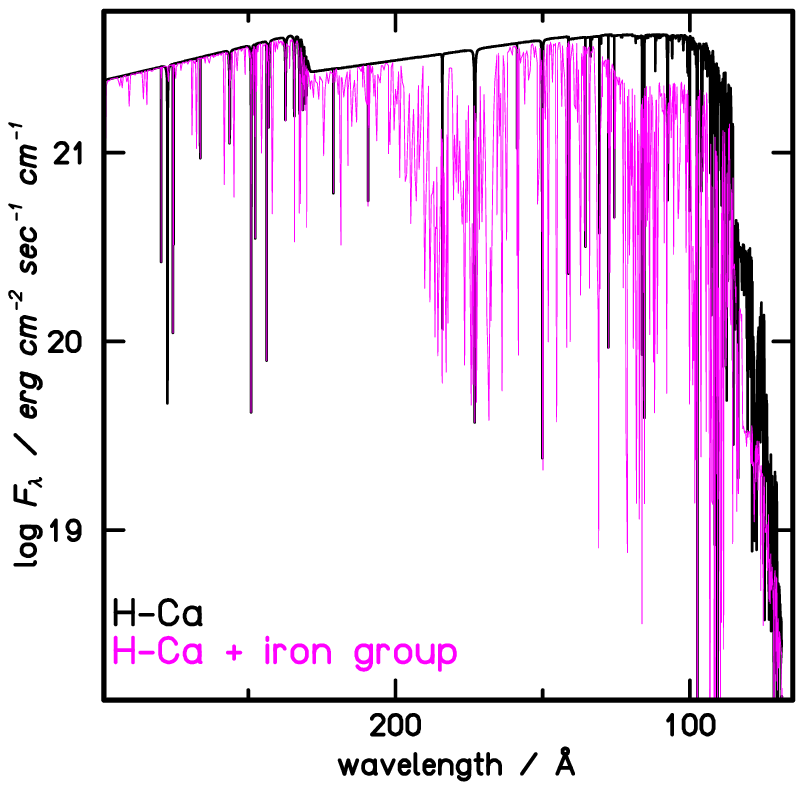}
\end{minipage}\hspace{5mm}
\begin{minipage}{4cm}
Figure 4. Comparison of two NLTE model atmosphere fluxes without
and with consideration of iron-group elements at solar abundances ($T_{\rm eff} = 155$\,kK, $\log g = 6.5$).
The additional iron-group line blanketing has a strong influence on the flux level at 
high energies close to the maximum flux level
\end{minipage}\vspace{-4mm}

\section{Conclusions and future work}
Emergent fluxes calculated from NLTE model atmospheres which
include iron-group line blanketing show a drastic decrease of  
the flux level at high energies.
For a reliable analysis of UV/EUV and X-ray spectra of central stars of planetary nebulae, or the
calculation of ionizing spectra from these (e.g.\,used as input
for photoionization models) the consideration
of all elements from hydrogen up to the iron group is mandatory.
A detailed consideration of the metal-line blanketing with all available lines
has an important influence on the spectrum.

{\sc PRO2} is permanently updated in order to calculate state-of-the-art models
for the analysis of the available spectra. This includes in the near future
spherical geometry, element diffusion, and polarized radiation transfer.

A new grid of model fluxes which includes a detailed line blanketing by Ca and by the iron group
will be soon available on the WWW.

\section*{Acknowledgment}
This research is supported by the DLR under 50\,OR\,9705\,5.

\end{document}